\documentclass[twocolumn,showpacs,preprintnumbers,amsmath,amssymb,superscriptaddress]{revtex4}
\usepackage{graphicx}% Include figure files
\usepackage{bm}% bold math

\begin{document}

\title{A new platform for topological quantum phenomena: Topological Insulator states in thermoelectric Heusler-related ternary compounds }% Force line breaks with \\

\author{H. Lin}
\affiliation{Joseph Henry Laboratories of Physics, Princeton University, Princeton, New Jersey 08544, USA}
\affiliation{Department of Physics, Northeastern University, Boston, Massachusetts 02115, USA}

\author{L.A. Wray}
\affiliation{Joseph Henry Laboratories of Physics, Princeton University, Princeton, New Jersey 08544, USA}
\author{Y. Xia}
\affiliation{Joseph Henry Laboratories of Physics, Princeton University, Princeton, New Jersey 08544, USA}
\author{S. Jia}
\affiliation{Department of Chemistry, Princeton University, Princeton, New Jersey 08544, USA,}
\author{R.J. Cava}
\affiliation{Department of Chemistry, Princeton University, Princeton, New Jersey 08544, USA,}
\author{A. Bansil}
\affiliation{Department of Physics, Northeastern University, Boston, Massachusetts 02115, USA}
\author{M.Z. Hasan}
\affiliation{Joseph Henry Laboratories of Physics, Princeton University, Princeton, New Jersey 08544, USA}
\affiliation{Princeton Center for Complex Materials, Princeton University, Princeton, New Jersey 08544, USA}
\affiliation{Princeton Institute for Science and Technology of Advanced Materials. PRISM, Princeton University, Princeton, New Jersey 08544, USA}
\email{mzhasan@princeton.edu}

\date{\today}% It is always \today, today,
             %  but any date may be explicitly specified

\begin{abstract}
Topological insulators realize a novel state of quantum matter that are distinguished by topological invariants of bulk band structure rather than spontaneously broken symmetries. A number of exotic quantum phenomena have been predicted to exist in multiply-connected geometries which require an enormous amount of materials flexibility. We have extended our previous search for TI materials from binary (Bi$_2$X$_3$ series, Xia \emph{et.al.}, Nature Phys. \textbf{5}, 398 (2009)) to the thermoelectric ternary compounds. We discover that the distorted LuPtSb is the first ternary compound of the ``$MM'X$" series harboring a 3D topological insulator state with Z$_2$=-1 whereas TiNiSn is trivial. We also show that the half-Heusler LuPtSb-type series is a natural platform that hosts a range of candidate compounds, alloys and artificial heterostructures (quantum-wells). We also discovered several different paradigms of trivial and non-trivial topological ordering in this class, including an intrinsically metallic nontrivial topological state in YAuPb. Some of these materials are grown (results will be reported separately).
\end{abstract}

\maketitle

Topological insulators (TI)\cite{Kane1st,FuKane,DavidNat1,TIbasic,BernevigSciHgTe,KonigSci, MooreandBal} realize a novel state of quantum matter that are distinguished by topological invariants of bulk band structure rather than spontaneously broken symmetries. Its material realization in 2D artificial HgTe-quantum wells \cite{BernevigSciHgTe,KonigSci} and 3D Bi-based binary compounds \cite{FuKane, DavidNat1,DavidTunable,DavidScience,MatthewNatPhys, Noh,Roushan, ZhangPred,ChenBiTe,BiTeSbTe, WrayCuBiSe} led to a surge of interest in discovering novel topological physics in world-wide condensed matter community. A number of exotic quantum phenomena have been predicted to exist in multiply-connected geometries \cite{Majorana, ZhangDyon, KaneSCproximity, KaneDevice, cenke, palee, dhlee} which require an enormous amount of materials flexibility. Given the right materials, these topological properties naturally open a window to new realms of spintronics and quantum computing. Just as the majority of normal metals are neither good superconductors nor strongly magnetic, there is no inherent likelihood that the band structure of topological insulators known so far will be suitable for the realization of any of the novel physical states or devices that have been predicted for quantum computing or spintronics. We need to expand our search.
We have extended our previous search for TI materials from binary (Bi$_2$X$_3$ series, Xia \emph{et.al.}, Nature Phys. \textbf{5}, 398 (2009)) to the thermoelectric ternary compounds. We discover that the distorted LuPtSb is the first ternary compound of the ``$MM'X$" series harboring a 3D topological insulator state with Z$_2$=-1. We also show that the half-Heusler LuPtSb-type series is a natural platform that hosts a range of candidate compounds, alloys and artificial heterostructures (quantum-wells). We also discovered several different paradigms of trivial and non-trivial topological ordering in this class, including an intrinsically metallic nontrivial topological state in YAuPb.
In this Letter, we use first principles theoretical calculations to show that distorted LuPtSb is the first ternary compound expected to realize a three dimensional topological insulator state. 

%The surface was simulated by placing a slab of [xx] layers in vacuum.
%%Local orbitals with $p_{1/2}$ radial wavefunctions were used to reduce the dependence on the atomic spheres and the number of scalar-relativistic eigenstates.

The crystal lattice of half-Heusler ternary compounds is described by the space group $F\bar{4}3m$, with the atomic arrangement shown in Fig.~\ref{fig:sketch}a. Crystalline compounds of this structure are very common, and are customarily assigned the chemical formula ``$MM'X$", where $M$, $M'$, and $X$ atoms occupy the Wyckoff 4b, 4c, and 4a positions respectively \cite{Wyckoff}. The $M'$ and $X$ atoms form a zincblende lattice, when taken alone. Atoms of $M$ fill empty space within the zincblende structure and form NaCl-like ionic pairs together with $X$ atoms. These materials closely resemble the previously known topologically nontrivial compound HgTe \cite{BernevigSciHgTe,KonigSci}, which achieves topological order with a zincblende structure, and can be considered to belong to the class of half-Heusler TIs if one assigns a vacancy to the $M$ site ($M$= void, denoted by []) and sets $M'$=Hg and $X$=Te. There is no spatial inversion symmetry in zincblende and half-Heusler structures.

\begin{figure*}
\includegraphics[width=13cm]{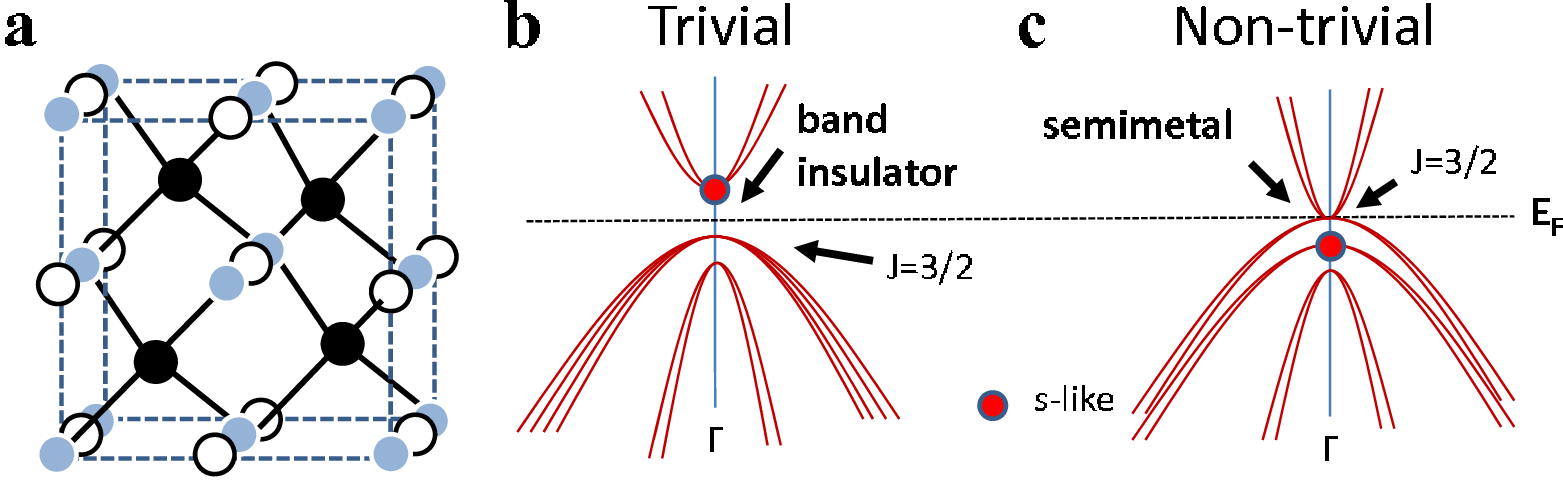}
\caption{\label{fig:sketch}
\textbf{Half-Heusler crystal structure and band inversion}: \textbf{a} The crystal structure of half-Heusler compounds $MM'X$. $M$, $M'$, and $X$ is denoted by open, black, and gray circles. The zincblende structure formed by $M'$ and $X$ is emphasized by black lines. Diagrams in \textbf{b}-\textbf{c} illustrate band structures near the $\Gamma$-point for trivial and non-trivial cases, respectively. Red dots denote the s-like orbitals at $\Gamma$-point. Band inversion occurs in the non-trival case where the s-like orbitals at the $\Gamma$-point are below the four-fold degenerate $j=3/2$ bands.
}
\end{figure*}

Hg$_{1-x}$Cd$_x$Te is a family of strong spin orbit 3D materials (Dornhaus and Numtz, 1983)\cite{HgTe}. In the electronic structure of CdTe the conduction band edge states have an \emph{s} like symmetry, while the valence band edge states have a \emph{p} like symmetry. In HgTe, the \emph{p} levels rise above the \emph{s} levels, leading to an inverted band structure. Due to this similarity, it is instructive to begin our discussion of half-Heusler band structure topology by examining the nontrivial topological nature of well studied 3D-HgTe. While the related material, CdTe (half-Heusler []CdTe) is a typical band insulator, HgTe is topologically nontrivial due to the inversion of two groups of bands with respective \emph{s}- and \emph{p}- orbital symmetry at the $\Gamma$-point. The different ordering of band symmetries at the $\Gamma$-point for trivial and non-trivial cases are summarized in Fig.~\ref{fig:sketch}b-c, with red dots labeling the bands with large \emph{s}-orbital occupancy. In the trivial case, the \emph{s}-like orbital is above the band gap, and in the nontrivial case, the \emph{s}-like orbital at the $\Gamma$-point is below the four-fold degenerate $j=3/2$ bands \cite{BernevigSciHgTe}\cite{FuKane}. This band inversion only occurs near the $\Gamma$-point and is absent at other special time reversal invariant points. We will show that ternary half-Heusler compounds also exhibit these two topologically distinct classes, distinguished in the same way through orbital symmetries on the $X$-sublattice.

The calculated (DFT-GGA) band structures of three characteristic half-Heusler materials, LuPtSb, YAuPb, and TiNiSn along high symmetry lines in the Brillouin zone are presented in Fig.~\ref{fig:bulkbands}. A common feature of these materials is that the top of the valence band is located at the $\Gamma$-point. Away from $\Gamma$, the Fermi level is completely gapped with the exception of electron pockets found at the X-point in YAuPb, where a direct gap is still present between the conduction and valence bands. Therefore, the topological properties can be determined from observations of band structure only near the $\Gamma$-point. Other materials with similar properties are addressed in the supplementary information.

Confining our view to band structure very close to the Fermi level (Fig.~\ref{fig:bulkbands}a(inset)), we find that the orbital angular momentum symmetries of these ternary half-Heusler compounds are identical to those defining low energy properties in HgTe and CdTe. For LuPtSb, two upward-concave bands and two downward-concave bands are degenerate at the $\Gamma$ point. The Fermi energy should lie exactly at this degenerate point for undoped samples. These four-fold degenerate states at the $\Gamma$-point have \emph{p}-type orbital symmetry with a total angular momentum eigenvalue of $j=3/2$, and lie above a two-fold degenerate s-like state, representing an inversion relative to the natural order of \emph{s}- and \emph{p}-type orbital derived band structure. Away from the $\Gamma$-point, the upward dispersing bands gain significant \emph{s}-like character due to orbital hybridization. The same band inversion also occurs in YAuPb, but low energy properties in YAuPb are complicated by another set of conduction bands that comes below the Fermi energy to form electron pockets near $X$. Thus, by analogy with HgTe, we expect YAuPb to be a topologically nontrivial metal (or ``topological metal"), as is the case with elemental antimony \cite{DavidScience}, and its conductivity could probably be manipulated through alloying, just as antimony was alloyed with bismuth to discover the first known example of a three dimensional TI \cite{DavidNat1} (Bi$_{1-x}$Sb$_{x}$). The s-type bands in TiNiSn are above $E_F$ and very high in energy (not shown), above the $j=3/2$ bands, meaning that its band structure lacks the $s/p$ inversion that leads to strong topological order. As we will establish more rigorously below, these symmetries correctly indicate that TiNiSn is topologically trivial while LuPtSb and YAuPb are topologically nontrivial.

\begin{figure*}
\includegraphics[width=13cm]{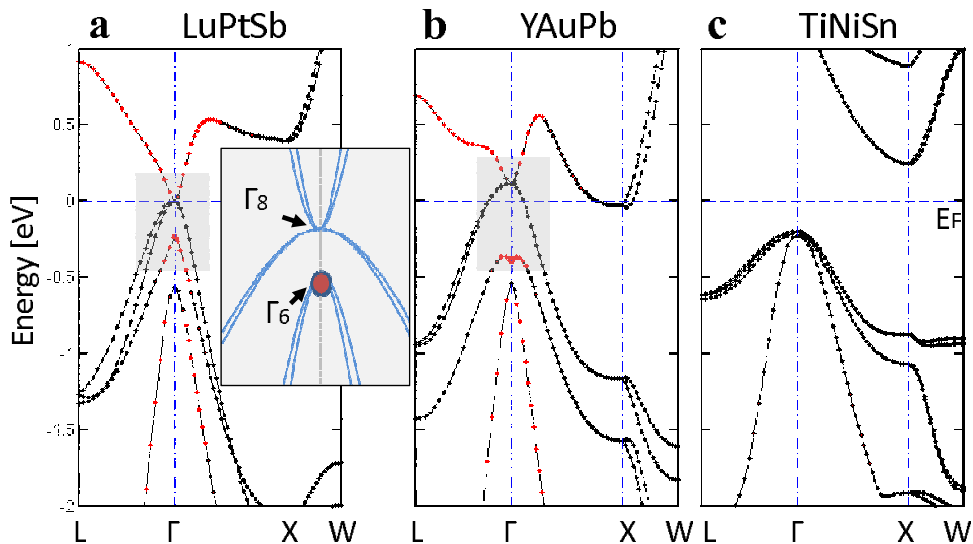}
\caption{\label{fig:bulkbands}
\textbf{Half-Heusler band structures}: Bulk band structures of (\textbf{a}) LuPtSb, (\textbf{b}) YAuPb, and (\textbf{c}) TiNiSn. The size of red data points is proportional to the probability of s-orbital occupation on the anion atom (``$X$"). An inset in panel \textbf{a} highlights inverted band symmetries associated with topological order, corresponding to the the diagram in Fig. 1c.
}
\end{figure*}

In terms of the symmetry notation that has been applied to HgTe in previous literature, our calculations show that bands near $E_F$ at the $\Gamma$-point possess $\Gamma_6$ (2-fold degenerate), $\Gamma_7$ (2-fold degenerate), and $\Gamma_8$ (4-fold degenerate) symmetry in all of these compounds. Both LuPtSb and HgTe have $\Gamma_8$ states at $E_F$. The $\Gamma_6$ symmetry bands are below $\Gamma_8$ and occupied, providing the band inversion that distinguishes topological order. Although the order of $\Gamma_7$ and $\Gamma_6$ is different, it is not relevant to the topological nature since both are occupied. In the band structures of topologically trivial band insulators TiNiSn and CdTe, the $\Gamma_6$ states are above the downward dispersing valence bands. These non-inverted $\Gamma_6$ states lie above $E_F$ and are unoccupied. It is due to the occupancy of $\Gamma_6$ bands at the $\Gamma$-point in LuPtSb that the $Z_2$ topological invariant picks up an extra factor of -1 when compared to TiNiSn.

Therefore, the half-Heusler ternary compound YPtSb with expanded (over LuPtSb) lattice constants is proven to be a topological semimetal, sharing a common $Z_2$ topological invariant with TIs. When we replace Y by Lu to get LuPtSb the low energy band topology is unchanged, because Y$^{+3}$ and Lu$^{+3}$ are both first column transition metals differing by the presence of 5$s^2$4$d^{10}$5$p^6$4$f^{14}$ filled shells. The additional occupied states are complete shells and the $Z_2$ invariant cannot change. Finally, we continuously shrink the lattice until the experimental lattice constants of LuPtSb are reached. In the whole process, the $\Gamma_6/\Gamma_8$ band inversion at the $\Gamma$-point persists and the low energy Hamiltonian undergoes an adiabatic transformation without phase transition. Therefore, we have proven that LuPtSb belongs to the nontrivial $Z_2$=-1 class. 

\begin{figure*}
\includegraphics[width=12cm]{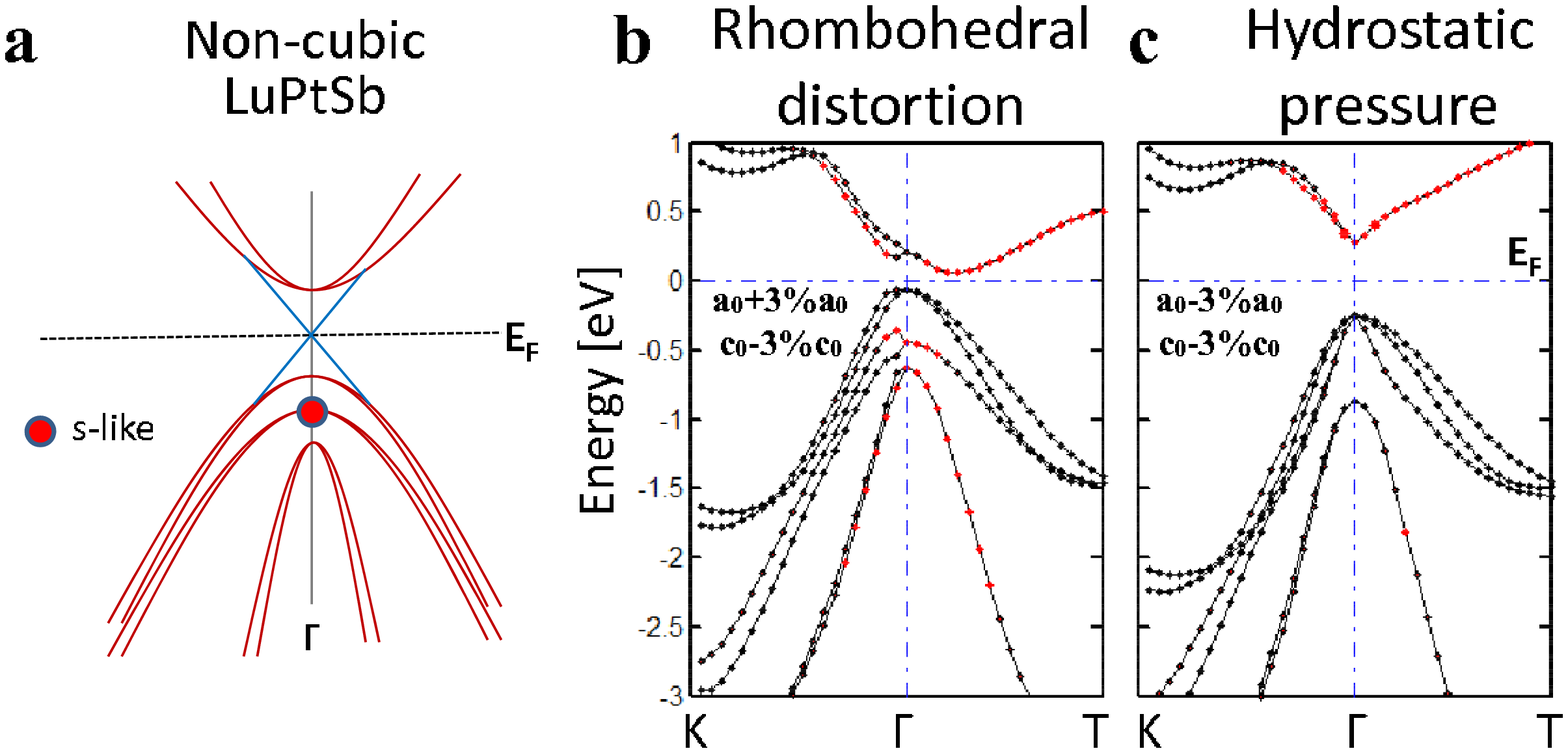}% Here is how to import EPS art
\caption{\label{fig:distortion}
\textbf{The topological insulator state with a single-Dirac-cone}: (a) A sketch of band structure near the $\Gamma$-point for topologically non-trivial LuPtSb with a lattice distortion. The s-like $\Gamma_6$ states are marked with a red dot. Lattice distortion causes a gap to open at $E_F$, resulting in a topological insulator state. For such a topological insulator, the surface bands will span the bulk band gap and should have an odd number of Dirac cones, resembling the dispersion plotted with blue lines. (b) Band structures of rhombohedral distorted LuPtSb with lattice constants $a=a_0+3\%a_0$ and $c=c_0-3\%c_0$ where $a_0$ and $c_0$ correspond to the experimental lattice. (c) Band structures of LuPtSb under hydrostatic pressure where $a=a_0-3\%a_0$ and $c=c_0-3\%c_0$.
}
\end{figure*}

The fact that none of the topologically nontrivial materials discussed in this work (3D-HgTe, LuPtSb, YAuPb) are naturally insulating known to be due to the $\Gamma$-point degeneracy of positive- and negative-mass bands with $\Gamma_8$ symmetry, which results from the crystal symmetry group. Distortion of the crystal lattice through static pressure or finite size effects can lift the degeneracy and open a gap at $E_F$, causing a direct gap between the topologically inverted valence and conduction bands and resulting in the appearance of topologically defined surface states. Fu and Kane have previously discussed this possibility for 3D-HgTe\cite{FuKane}. In Fig.~\ref{fig:bulkbands}b, we demonstrate that a perturbative rhombohedral lattice distortion caused by uniaxial pressure will lift the band structure degeneracy in LuPtSb, resulting in a fully gapped topological insulator state. With reference to the standard axial labeling convention for half-Heusler compounds, a 3\% compression is applied along the [111] direction and a 3\% expansion in the plane perpendicular to [111]. As has been observed when distortion is applied to the zincblende structured binary compound HgTe, bulk band structure of the distorted half-Heusler LuPtSb is fully gapped and realizes strong Z$_2$ topological order.

Conversely, the topological band inversion in LuPtSb can be removed altogether by uniformly decreasing all lattice constants, demonstrating the sensitive chemical tunability in this TI class (Fig.~\ref{fig:bulkbands}c). After a 3\% reduction in all lattice parameters, the s-like $\Gamma_6$ symmetry bands are observed to rise above the $\Gamma_8$ bands, removing band inversion at the $\Gamma$-point and resulting in a topologically trivial band insulator state.

In conclusion, we have discovered a new class of topological insulator materials realized by distorted half-Heusler compounds, and demonstrated band structure topology calculations for representative topologically non-trivial semimetal (LuPtSb) and nontrivial metallic (YAuPb) undistorted materials. Bulk band structure in LuPtSb is shown to be characterized by Z$_2$=-1. A large number of topological compounds exist (list to be discussed elsewhere) in this ternary material class, making it a versatile platform for exploring many different device configurations for realizing topological quantum phenomena not accessible in the binary topological insulator classes such as the Bi$_2$Se$_3$ series discovered previously.

M.Z.H. acknowledges discussions with C.L. Kane and B.A. Bernevig and support from U.S.DOE and A.P. Sloan Research Fellowship. H.L. acknowledges support from Northeastern and Princeton University. R.J.C. acknowledges discussions and long-standing collaborations with C. Felser and T. Kilmczuk on thermoelectric and superconducting-Heusler phases and with Y.S. Hor on Ternary-topological-materials.

\textbf{Methods summary:}
First-principles band calculations were performed with the linear augmented-plane-wave (LAPW) method using the WIEN2K package \cite{wien2k} in the framework of density functional theory (DFT). The generalized gradient approximation (GGA) of Perdew, Burke, and Ernzerhof \cite{PBE96} was used to describe the exchange-correlation potential. Spin orbital coupling (SOC) was included as a second variational step using a basis of scalar-relativistic eigenfunctions.

%\bibliography{draft}% Produces the bibliography via BibTeX.

\newpage

\begin{figure*}
\includegraphics[width=13cm]{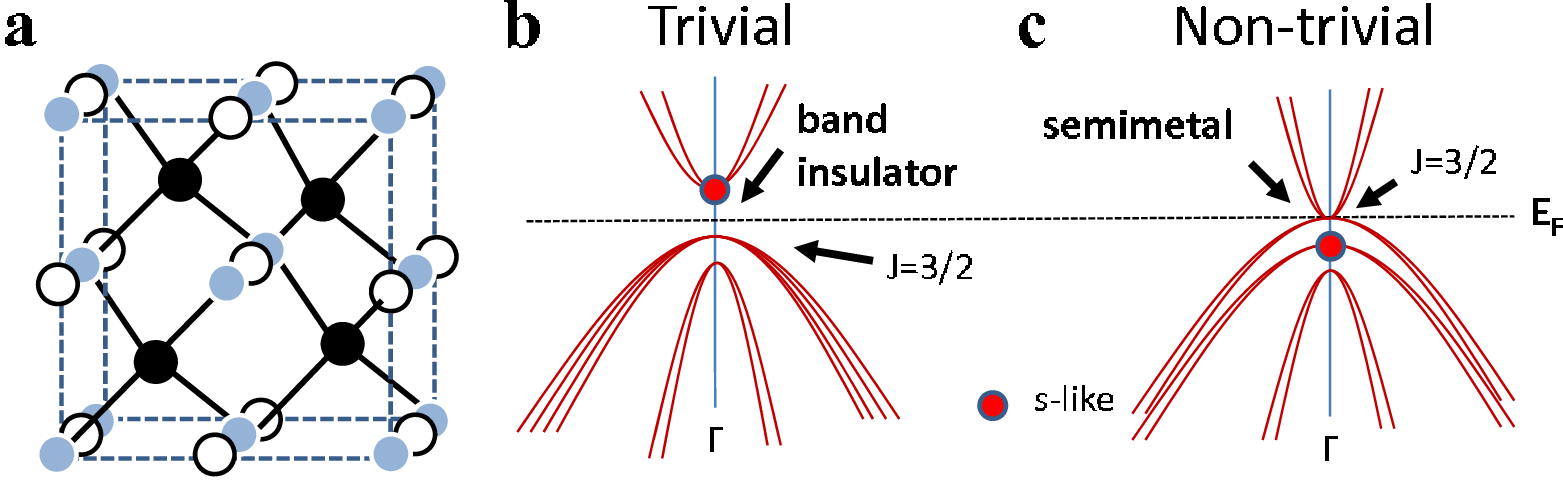}
\caption{\label{fig:sketch}
\textbf{Half-Heusler crystal structure and band inversion}: \textbf{a} The crystal structure of half-Heusler compounds $MM'X$. $M$, $M'$, and $X$ is denoted by open, black, and gray circles. The zincblende structure formed by $M'$ and $X$ is emphasized by black lines. Diagrams in \textbf{b}-\textbf{c} illustrate band structures near the $\Gamma$-point for trivial and non-trivial cases, respectively. Red dots denote the s-like orbitals at $\Gamma$-point. Band inversion occurs in the non-trival case where the s-like orbitals at the $\Gamma$-point are below the four-fold degenerate $j=3/2$ bands.
}
\end{figure*}

\begin{figure*}
\includegraphics[width=13cm]{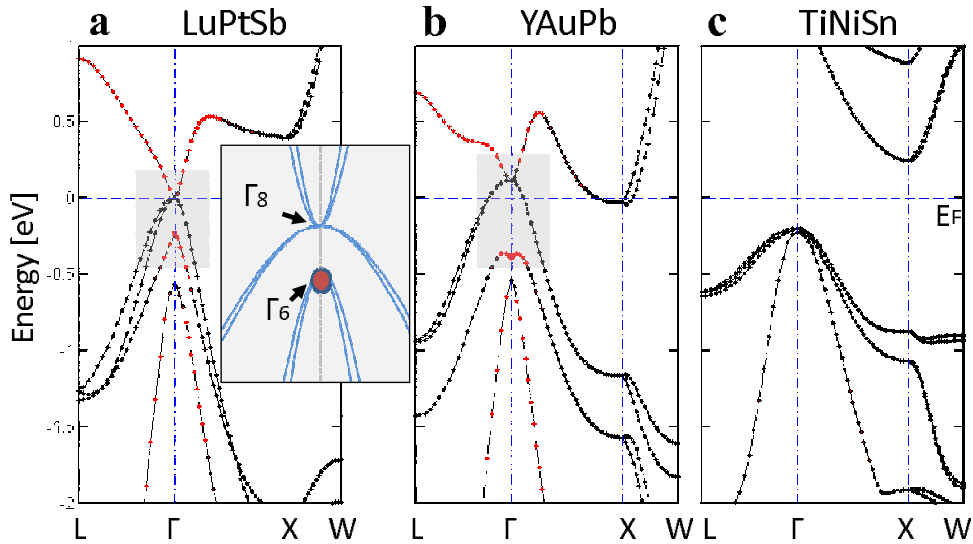}
\caption{\label{fig:bulkbands}
\textbf{Half-Heusler band structures}: Bulk band structures of (\textbf{a}) LuPtSb, (\textbf{b}) YAuPb, and (\textbf{c}) TiNiSn. The size of red data points is proportional to the probability of s-orbital occupation on the anion atom (``$X$"). An inset in panel \textbf{a} highlights inverted band symmetries associated with topological order, corresponding to the the diagram in Fig. 1c.
}
\end{figure*}

\begin{figure*}
\includegraphics[width=12cm]{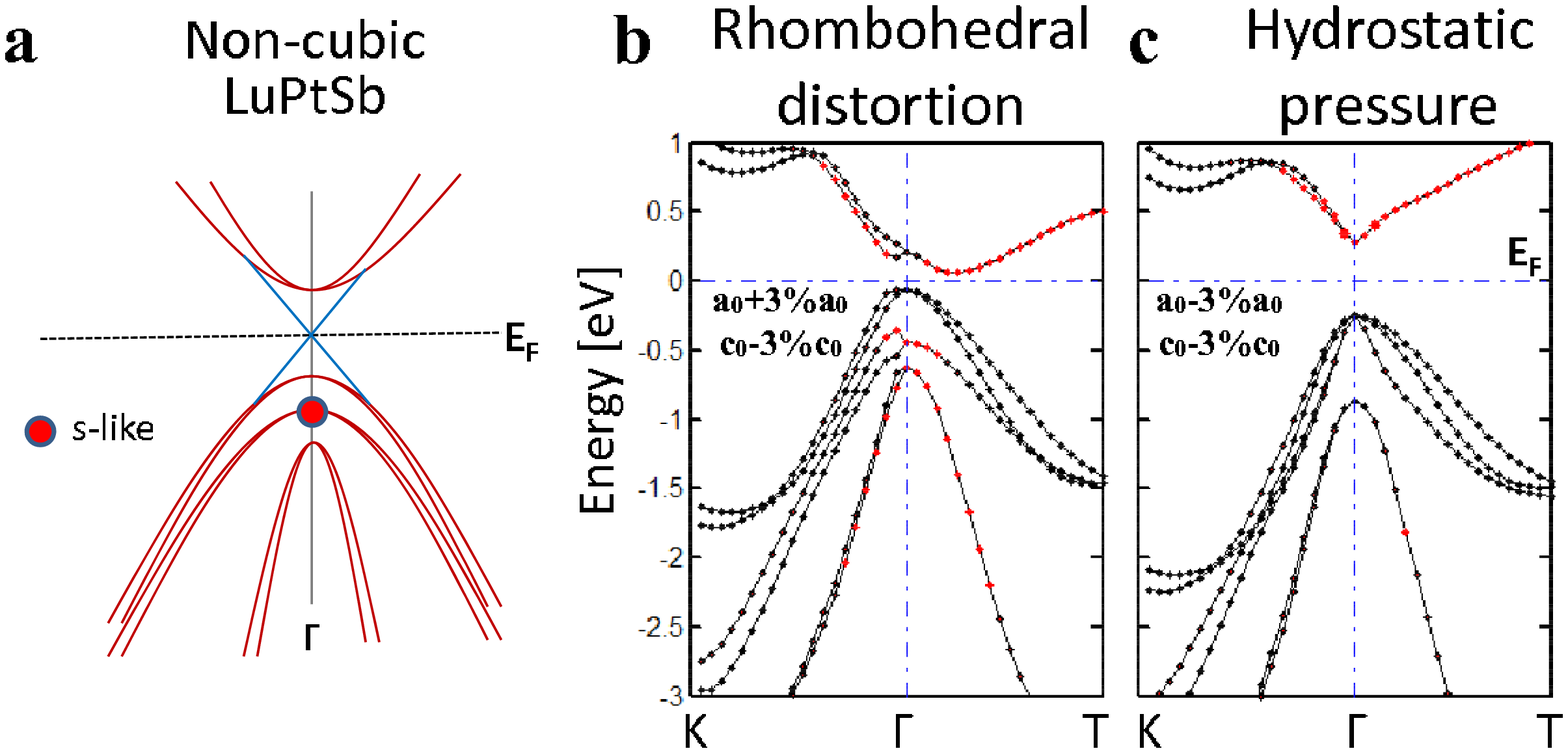}% Here is how to import EPS art
\caption{\label{fig:distortion}
\textbf{The topological insulator state with a single-Dirac-cone}: (a) A sketch of band structure near the $\Gamma$-point for topologically non-trivial LuPtSb with a lattice distortion. The s-like $\Gamma_6$ states are marked with a red dot. Lattice distortion causes a gap to open at $E_F$, resulting in a topological insulator state. For such a topological insulator, the surface bands will span the bulk band gap and should have an odd number of Dirac cones, resembling the dispersion plotted with blue lines. (b) Band structures of rhombohedral distorted LuPtSb with lattice constants $a=a_0+3\%a_0$ and $c=c_0-3\%c_0$ where $a_0$ and $c_0$ correspond to the experimental lattice. (c) Band structures of LuPtSb under hydrostatic pressure where $a=a_0-3\%a_0$ and $c=c_0-3\%c_0$.
}
\end{figure*}

\end{document}